\documentclass[12pt]{article}
\usepackage{bm,amssymb}
\usepackage{hyperref}
\usepackage[pdftex]{graphicx}
\usepackage{graphics,graphicx}
\usepackage{color}
\usepackage{graphicx}
\usepackage[format = plain,labelfont = bf,up, textfont = normal , up, justification =raggedright, singlelinecheck =false]{caption}
\usepackage{subcaption}
\usepackage{tabularx}
\usepackage{amsmath}

\usepackage{graphicx}
\usepackage[utf8x]{inputenc}
\usepackage{color,soul}
\usepackage{amsmath}
\usepackage{braket}
\usepackage{latexsym}
\usepackage{amssymb}
\usepackage{amsthm}
\usepackage{bm}
\usepackage{graphics,epstopdf}
\usepackage{color}\usepackage{amsmath}
\usepackage{enumitem}
\usepackage{fmtcount}
\usepackage{booktabs}

\usepackage{epsfig}

\theoremstyle{plain}

\def\bea{\begin{eqnarray}}
\def\eea{\end{eqnarray}}
\def\ba{\begin{array}}
\def\ea{\end{array}}

\def\ket{\rangle}

\def\beq{\begin{equation}}
\def\eeq{\end{equation}}

\begin{document}
\title{Signaling versus distinguishing different preparations of same pure quantum state}


\author{Chirag Srivastava\footnote{E-mail: \href{chiragsrivastava@hri.res.in}{chiragsrivastava@hri.res.in}},\, Sreetama Das\footnote{E-mail: \href{sreetamadas@hri.res.in}{sreetamadas@hri.res.in}},\, Aditi Sen(De)\footnote{E-mail: \href{aditi@hri.res.in}{aditi@hri.res.in}}, and Ujjwal Sen\footnote{E-mail: \href{ujjwal@hri.res.in}{ujjwal@hri.res.in}}}

\date{}
\maketitle
\vspace{-1cm}
\begin{center}
Harish-Chandra Research Institute, HBNI,\\ Chhatnag Road, Jhunsi, Allahabad 211 019, India
\end{center}

\begin{abstract}
Different ensembles of quantum states can have the same average nonpure state. Distinguishing between such constructions, via different mixing procedures of the same nonpure quantum state, is known to entail signaling. In parallel, different superpositions of pure quantum states can lead to the same pure  state. We show that the possibility of distinguishing between such preparations, via different interferometric setups leading to the same pure quantum state, also implies signaling. The implication holds irrespective of whether the distinguishing procedure is deterministic or probabilistic.
\end{abstract}

\section{Introduction}
 The state of a quantum system, if it is not pure, always has an infinite number of decompositions into ensembles. It is postulated within quantum mechanics 
that these decompositions represent the same physical situation. 
Indeed, it is known that if different decompositions of the same mixed quantum state are distinguishable, 
it will result in instantaneous communication of information between two spatially separated parties, who can in principle be even space-like separated
\cite{ghirardi,gisin89,wooters}.
In general, it has been found that ``tweaking" quantum evolutions, i.e., considering non-quantum evolutions, typically result 
in signaling
\cite{gisin89,gisin90,gisin95,horodecki-ra,svet,sir,ferraro}. 
It is interesting to note that the no-signaling constraint has been used
with success in several areas. In particular, it has been used to obtain
the optimal approximate quantum cloning fidelity \cite{gisin98,
bruss}, and fidelity for optimal quantum state discrimination \cite{bae}. Also, there has been an important set of works that
considered the security in bit commitment protocols based on the no-signaling principle \cite{kent1}. Cryptographic protocols have also been considered where the eavesdropper
is restrained not by quantum mechanics but by the weaker condition of
no-signaling \cite{kent2}. It is interesting to note here that the no-signaling condition has recently been generalized, leading to interesting consequences \cite{pawlowski}.

\begin{figure}[h]
\begin{center}
\includegraphics[width = 0.5\textwidth]{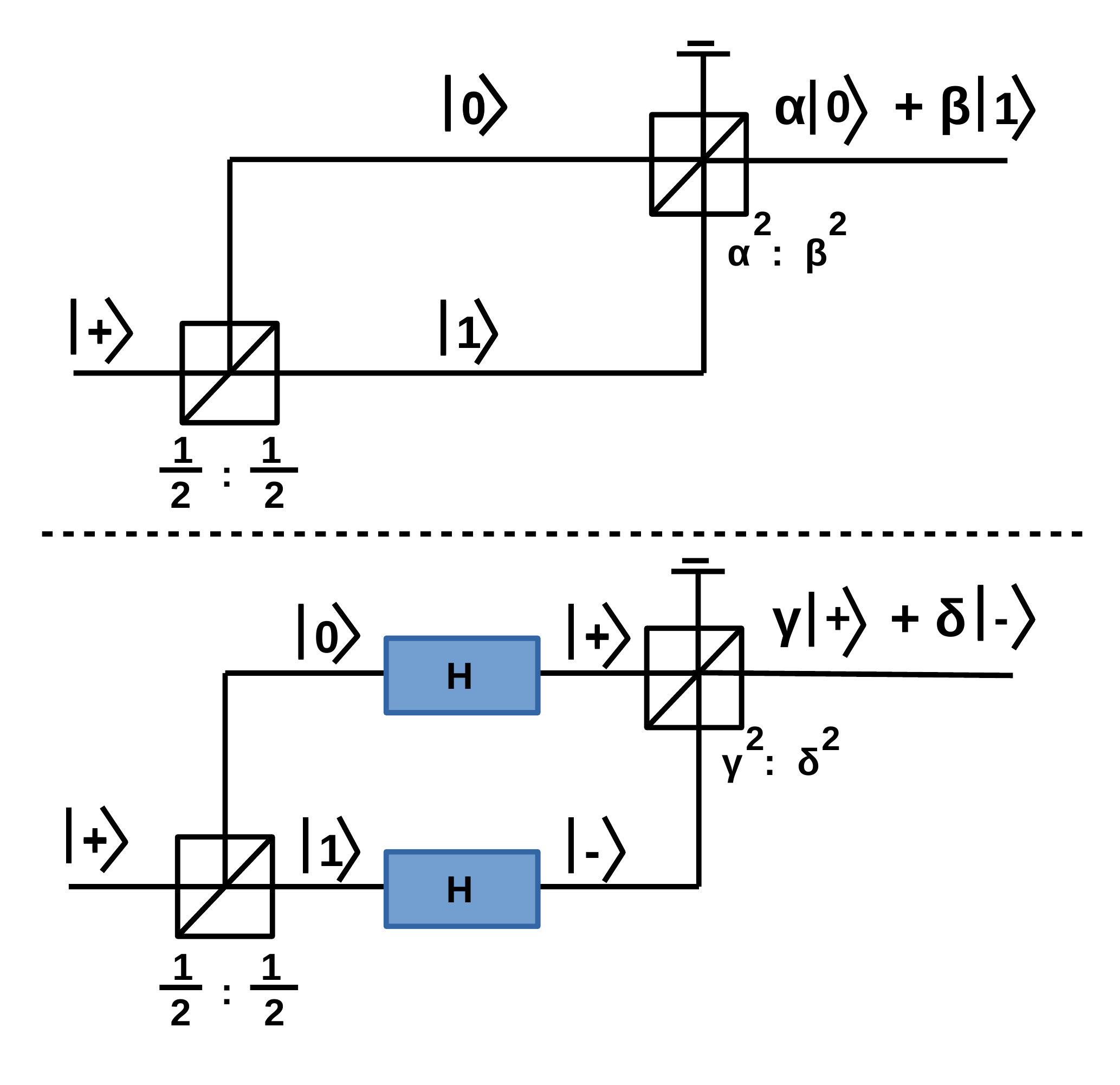} 
\caption{Two preparation procedures of the same state. We consider a photonic set-up, where we denote the horizontal and vertical polarization states as $|0\ket$ and $|1\ket$ respectively, and
 $|+\ket=\frac{1}{\sqrt{2}}(|0\ket+|1\ket)$. The square boxes represent polarization beam splitters, with their splitting ratios being indicated below them in the figure. The two panels correspond to the two preparation procedures. In both the panels, the input is from the left, and is the state $|+\ket$. In the upper panel, the vertical final output is discarded, while the state in the horizontal final  output channel is $|\psi_{\alpha \beta}\ket=\alpha|0\ket+\beta|1\ket$, where \(\alpha\) and \(\beta\) are complex numbers, and \(|\alpha|^2+|\beta|^2=1\). In the lower panel, the set-up is otherwise the same, but there are two Hadamard gates inserted, denoted in the figure as rectangular boxes with ``H" written on them. Again, the vertical final output is disposed of, and the horizontal final output is in the state 
 $|\psi_{\gamma \delta}\ket=\gamma|+\ket+\delta|-\ket$, where $\alpha,~\beta,~\gamma,~\delta$ are so chosen that $\alpha|0\ket+\beta|1\ket = \gamma|+\ket+\delta|-\ket$.
}
\label{fig.1}
\end{center}
\end{figure}

If the quantum state of a physical system is pure, it can be represented by superpositions over an infinite number of bases of the corresponding Hilbert space.  While different decompositions of the same mixed state can be perceived of as different mixing strategies during the preparation stage of the mixed state, different superpositions of the same pure state can be viewed as different interferometric-type setups during the preparation stage of the pure state. In this paper, we use different representations (superpositions) of the same pure state to indicate that they have been prepared differently, as shown in Fig.~\ref{fig.1}. 
Notice that these different preparation procedures of the same state, and an engagement in trying to distinguish between them is an inverse of the attempt to identify and correct errors in the same state in quantum error correction protocols \cite{shor}. Error correction codes do exist within the ambit of quantum mechanics.

Here we show that a possibility of discrimination of different preparation procedures of the same pure state, which is postulated in quantum mechanics to be impossible, will result in signaling.    


\section{The association}
\begin{figure}[h]
\begin{center}
\includegraphics[width = 0.5\textwidth]{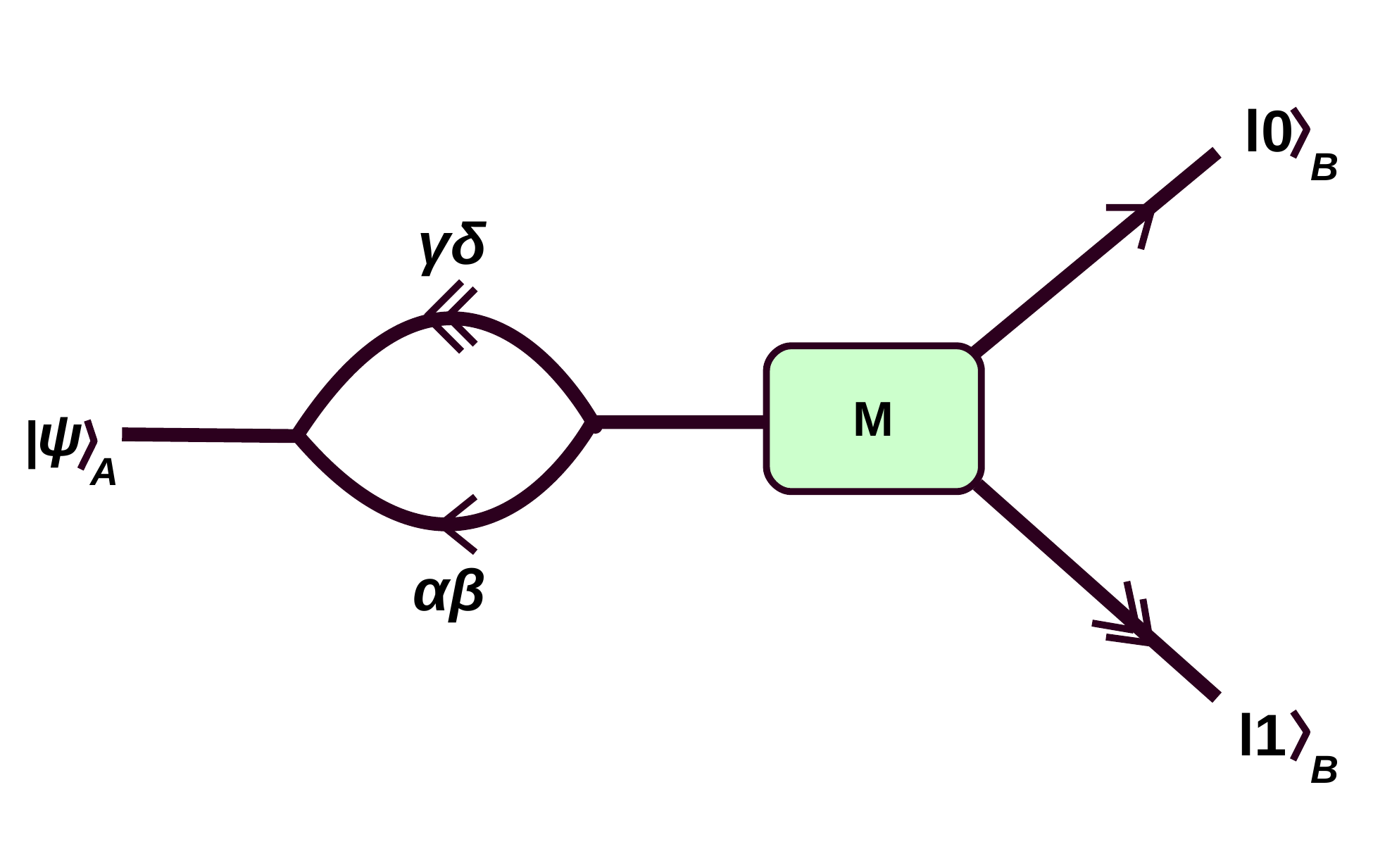} 
\caption{Schematic potential preparation procedure for the state $|\Phi\ket_{AB}$. The state, $|\Phi\ket_{AB}$, given in Eq. (\ref{Eq.1}), can in principle be prepared by a machine, {\bf M}, whose arms on the right represents a qubit, spanned by $|0\ket_B$ and $|1\ket_B$, and is in possession of Bob. The output on the left of {\bf M} are in possession of Alice. The lower arm on the left represents the preparation procedure of the quantum state $|\psi\ket$ via the method presented in upper panel in Fig. \ref{fig.1}. Parallely, the upper arm on the left represents that via the method presented in lower panel in Fig. \ref{fig.1}. Similar to that in, say type-I spontaneous parametric down conversion \cite{hong}, upper and lower arms respectively on the right and left are correlated, and coherently superposes with instances when the lower and upper arms fire respectively on the right and left.    }  
\label{fig.2}
\end{center}
\end{figure}

 Consider two parties, Alice $(A)$ and Bob $(B)$, sharing a pure state, 
\begin{equation}  
 |\Phi\rangle_{AB} = \frac{1}{\sqrt{2}} \left( |\psi_{\alpha\beta}\rangle_A|0\rangle_B  + |\psi_{\gamma\delta}\rangle_A|1\rangle_B \right),
 \label{Eq.1}
 \end{equation}
 of two spin-1/2 quantum particles, where 
\begin{equation}
|\psi_{\alpha\beta}\rangle =  \alpha|0\rangle + \beta |1\rangle        
       \end{equation}
 is an expansion of the pure spin-1/2 quantum state \(|\psi\rangle_{A}\) in the 
\(\sigma_z\) basis, and 
\begin{equation}
|\psi_{\gamma\delta}\rangle = \gamma |+\rangle + \delta |-\rangle 
\end{equation}
 is an expansion of the same state \(|\psi\rangle_A\) in the 
\(\sigma_x\) basis. The two different representations, \(|\psi_{\alpha\beta}\ket\) and \(|\psi_{\gamma\delta}\ket\), are used to imply  that they are prepared differently, as shown in Fig. \ref{fig.1}. 
\(|0\rangle\) and \(|1\rangle\) of Bob's system are again eigenstates of \(\sigma_z\).
All kets and bras in this manuscript are normalized to unity, unless stated otherwise. 
The state, \(|\Phi\rangle_{AB}\), is equivalent to the product (unentangled) state \(|\psi\rangle_A \otimes |+\rangle_B\). A potential preparation schematic is presented in Fig.~\ref{fig.2}.    

Let us now assume that Alice has a machine that distinguishes between the two preparation procedures, \(|\psi_{\alpha\beta}\rangle\) and \(|\psi_{\gamma\delta}\rangle\), 
of the same pure state \(|\psi\rangle_A\). Note that the evolution (action of the distinguishing machine) is not quantum. But it is assumed that after its application on a situation, it returns some situation of the same physical system. And these situations are still represented by quantum states. Alice and Bob are in two separate locations, say a distance $(d)$ apart, 
and have had their clocks syncronized. 
It has also been decided that if it rains 
at time, $t_0$, at Alice's location, 
Alice will apply her machine to her part of the shared state. She won't do anything if it doesn't 
rain. 
It takes at least a time $\frac{d}{c}$ for light to travel from Alice's location to Bob's, where $c$ is the speed of light in vacuum.  A measurement at Bob's end, say at $t_0+d/2c$, on his part of 
the shared state, will reveal whether or not Alice had applied her machine. Indeed, suppose that the distinguishing protocol runs by implementing the transformations
\begin{eqnarray}
|\psi_{\alpha\beta}\rangle_A |0\rangle_{A_1} &\rightarrow & c_{\alpha\beta}|\overline{\psi}_{\alpha\beta}\rangle_{AA_1}, \nonumber\\
|\psi_{\gamma\delta}\rangle_A |0\rangle_{A_1} &\rightarrow & c_{\gamma\delta} |\overline{\psi}_{\gamma\delta}\rangle_{AA_1}, \label{bhole-byom}
\end{eqnarray}
 where, for the success 
of the distinguishing protocol of Alice, we must have \(|\langle\overline{\psi}_{\alpha\beta}|\overline{\psi}_{\gamma\delta}\rangle| \ne 1\). \(c_{\alpha\beta}\) and 
\(c_{\gamma\delta}\) are arbitrary nonzero complex numbers. \(A_1\) is an auxiliary 
machine system at Alice's laboratory. 

We now assume that the machine is linear on the superposition 
\begin{equation}
\frac{1}{\sqrt{2}}\left(|\psi_{\alpha\beta}\rangle_A |0\rangle_{A_1} |0\rangle_B + 
                        |\psi_{\gamma\delta}\rangle_A |0\rangle_{A_1} |1\rangle_B \right), 
                        \label{12byom}
\end{equation}
although not necessarily on other superpositions.
%
The shared state of 
the two laboratories (consisting of \(A\), \(A_1\), and \(B\)), is exactly in the state in (\ref{12byom}), which, after the action of Alice's machine, given by (\ref{bhole-byom}), 
is the unnormalized state
\begin{equation}
\frac{1}{\sqrt{2}}\left(c_{\alpha\beta} |\overline{\psi}_{\alpha\beta}\rangle_{AA_1}|0\rangle_B  + c_{\gamma\delta} |\overline{\psi}_{\gamma\delta}\rangle_{AA_1}|1\rangle_B\right).
\end{equation} If 
\(|\langle\overline{\psi}_{\alpha\beta}|\overline{\psi}_{\gamma\delta}\rangle| \ne 1\), this state will be entangled \cite{4h}, in the $AA_1$:$B$ partition, i.e., it cannot be expressed as $\tilde{c}|\chi_{AA_1}\ket|\chi_B\ket$,  where $\tilde{c}$ is a complex number, and $|\chi_{AA_1}\ket$ and $|\chi_B\ket$ are vectors in the Hilbert spaces corresponding to $AA_1$ and $B$ respectively. Consequently, the local state at Bob's (as well as Alice's) end 
will possess a nonzero von Neumann entropy, where the von Neumann entropy of a density matrix $\sigma$ is given by $-\mbox{tr}(\sigma\log_2{\sigma})$. This is the situation if it rained at Alice's location at time $t_0$. If it didn't, then Alice will not apply the 
machine corresponding to Eq. (\ref{bhole-byom}), and consequently, the von Neumann entropy at Bob's (and Alice's) end will be vanishing, since Bob's state is then pure. To find the 
von Neumann entropy at Bob's end, Bob can perform a tomography of his 
part
 of the shared state, for which we need to assume that there existed several copies of the shared 
state, and Alice applied the transformation (\ref{bhole-byom}), separately, on all copies if it did rain at time $t_0$. The tomography was performed by Bob at $t_0+d/2c$. The value of the von Neumann
entropy of Bob's state indicates whether it rained or not at Alice's location. We therefore find that the assumption of the possibility of distinguishing between
 \(|\psi_{\alpha\beta}\rangle\) and \(|\psi_{\gamma\delta}\rangle\) leads to the signaling, i.e., superluminal transfer of classical information.

 
 
To reiterate, we have assumed (a) that a certain machine exists which can distinguish between two superpositions of the same state, and (b) that the machine is linear on a certain superposition. The conjunction of these two assumptions implies signaling. Assuming that no-signaling is a physical constraint that we wish to adhere to, (a) or (b) or both are wrong. It is true that the machine assumed in (a) is not unitary. Indeed, no unitary machine can distinguish between two different superpositions of the same state. Our argument is that a future non-quantum theory that may allow non-unitary operations, will still have to keep different superpositions of the same state as non-distinguishable, if it is linear on a particular superposition and if we want to adhere to no-signaling.

Suppose now that Alice is able to apply a transformation that only probabilistically (i.e., sometimes) provides her with the possibility of 
distinguishing between the two superpositions \(|\psi_{\alpha\beta}\rangle\) and \(|\psi_{\gamma\delta}\rangle\). If that probability is \(p\), then the 
transformation (\ref{bhole-byom}) needs to be replaced by one that implements 
\begin{eqnarray}
|\psi_{\alpha\beta}\rangle_A |0\rangle_{A_1} &\rightarrow & \tilde{c}_{\alpha\beta}\left(\sqrt{p}|\overline{\psi}_{\alpha\beta}\rangle_{AA_1} 
                             + \sqrt{1-p}|\overline{\chi}_{\alpha\beta}\rangle_{AA_1}\right) \equiv \tilde{c}_{\alpha \beta} |{\widetilde{\psi}}_{\alpha \beta}\rangle, \nonumber\\
|\psi_{\gamma\delta}\rangle_A |0\rangle_{A_1} &\rightarrow & \tilde{c}_{\gamma\delta}\left(\sqrt{p}|\overline{\psi}_{\gamma\delta}\rangle_{AA_1} 
                             + \sqrt{1-p}|\overline{\chi}_{\gamma\delta}\rangle_{AA_1}\right) \equiv \tilde{c}_{\gamma\delta} |{\widetilde{\psi}}_{\gamma\delta}\rangle,\nonumber \\ \label{tarak-byom}
\end{eqnarray}
where the projector onto the span of \(|\overline{\psi}_{\alpha\beta}\rangle\) and \(|\overline{\psi}_{\gamma\delta}\rangle\) is orthogonal to that onto the span of
\(|\overline{\chi}_{\alpha\beta}\rangle\) and \(|\overline{\chi}_{\gamma\delta}\rangle\). \(\tilde{c}_{\alpha\beta}\) and \(\tilde{c}_{\gamma\delta}\) are 
arbitrary nonzero complex numbers. 
The relative phases on the right-hand-sides of (\ref{tarak-byom}) have been absorbed in the definitions of 
\(|\overline{\chi}_{\alpha\beta}\rangle\) and \(|\overline{\chi}_{\gamma\delta}\rangle\). 
Note that the relevant states, \(|\overline{\psi}_{\alpha\beta}\rangle\) and \(|\overline{\psi}_{\gamma\delta}\rangle\), can be obtained with probability \(p\), assumed to be non-zero,
when a measurement, comprising of two projectors, viz. the projector onto the span of \(|\overline{\psi}_{\alpha\beta}\rangle\) and \(|\overline{\psi}_{\gamma\delta}\rangle\) and that onto the 
orthogonal complement of that span,
is performed.  
The transformation  (\ref{tarak-byom}) followed by the measurement is similar to the ``unitary-reduction'' machines that are used in probabilistic cloning of quantum states \cite{Duan-Guo-kamaal} and tasks like probabilistic teleportation, dense coding, and masking of quantum information \cite{tasks}. A difference of our ``probabilistic'' scheme with the unitary-reduction methods is that instead of a deterministic unitary evolution followed by a measurement, a deterministic non-unitary evolution (\ref{tarak-byom}) followed by a measurement  is considered here. Such probabilistic machines are different from probabilistic processes where 
different deterministic processes are applied probabilistically onto an input, e.g. in twirling operations (see e.g. \cite{ibm-huge}).
For the transformation (\ref{tarak-byom}) to offer a probabilistic protocol of distinguishing 
between \(|\psi_{\alpha\beta}\rangle\) and \(|\psi_{\gamma\delta}\rangle\), one must have 
\(|\langle \overline{\psi}_{\alpha\beta} | \overline{\psi}_{\gamma\delta} \rangle |\) strictly less than unity. In that case, the modulus of the inner product of the 
normalized portions in the 
right-hand-sides of the transformation (\ref{tarak-byom}) 
can be bounded above strictly by unity:
%
\begin{eqnarray}
|\langle \widetilde{\psi}_{\alpha \beta} | \widetilde{\psi}_{\gamma \delta} \rangle| &\leq & p |\langle \overline{\psi}_{\alpha\beta} | \overline{\psi}_{\gamma\delta} \rangle | + (1-p) |\langle \overline{\chi}_{\alpha\beta} | \overline{\chi}_{\gamma\delta} \rangle | \nonumber\\
&\leq & p |\langle \overline{\psi}_{\alpha\beta} | \overline{\psi}_{\gamma\delta} \rangle | + (1-p)<1.\nonumber
\end{eqnarray}
The first inequality follows from the fact that \(|a+b|\leq |a|+|b|\) for complex numbers \(a\) and \(b\). 
The second one follows by using \(|\langle \overline{\chi}_{\alpha \beta}| \overline{\chi}_{\gamma \delta}\rangle| \leq 1\). 
The third (strict) inequality follows from the fact that  as \(|\langle \overline{\psi}_{\alpha\beta} | \overline{\psi}_{\gamma\delta} \rangle | < 1\), and that $p \neq 0$. 
Even this possibility  of distinguishing between the states \(|\psi_{\alpha\beta}\rangle\) and \(|\psi_{\gamma\delta}\rangle\) with some probability leads 
to signaling. To see this, apply the transformation (\ref{tarak-byom}) to the state \(|\Phi\rangle_{AB}|0\rangle_{A_1}\) to find that although it is a product state, in the \(AA_1:B\) partition,
before application of the transformation, it becomes entangled after. 
Again, a linearity for the superposition in (\ref{12byom}) is assumed, although linearity for other superpositions may not hold.
The post-transformation state is given by 
\begin{equation}
\frac{1}{\sqrt{2}}\left(\tilde{c}_{\alpha\beta} |\widetilde{\psi}_{\alpha\beta}\rangle_{AA_1}|0\rangle_B  + \tilde{c}_{\gamma\delta} |\widetilde{\psi}_{\gamma\delta}\rangle_{AA_1}|1\rangle_B\right),
\end{equation}
which is entangled by virtue of the fact that \(|\langle \widetilde{\psi}_{\alpha \beta} | \widetilde{\psi}_{\gamma \delta} \rangle| < 1\), which in turn follows from the assumption that the states 
\(|\psi_{\alpha \beta}\rangle\) and \(|\psi_{\gamma \delta}\rangle\) are probabilistically distinguishable. 

\section{Conclusion}
 It is postulated in quantum mechanics that different preparation procedures of the same pure state cannot be discriminated. Assuming that a physical situation is always represented by a quantum state, it is shown that the existence of a machine that can discriminate between two preparation
procedures of the state leads to signaling. 

Note added: We note here that the results of Ref. \cite{pati'19} and of this manuscript were obtained in parallel and in knowledge of all authors of the two manuscripts. Although the results reported in the two manuscripts are similar, they 
use different proofs for
obtaining them. Specifically, the proof in \cite{pati'19} uses an initially entangled state shared between two parties, and tries to prove that different preparation procedures of same pure state leads to signaling. Whereas in the current manuscript, we use a state that is initially unentangled (product).
Additionally, and, we feel, importantly, the machine, in the current manuscript, that distinguishes between two preparation procedures of the same pure state uses an operation which is beyond quantum mechanics, and that leads to the distinguishing. The motivation here is to ask whether distinguishing different superpositions of the same pure
quantum state through a non-quantum process leads to signaling, just like it is already known that distinguishing different mixtures of the same mixed quantum state through a non-quantum process leads to signaling. The motivation is also to investigate whether tweaking quantum operations even slightly leads to violation of a physical principle like no-signaling, which
is independent of quantum theory.
Therefore, although the final results claimed in two manuscripts are similar, the perspectives are different. There is a probabilistic version of the distinguishing procedure that we consider in the current manuscript, which
is absent in \cite{pati'19}.

\section*{Acknowledgment.} 
We acknowledge discussions with Arun K. Pati. The research of SD was supported in part by the INFOSYS scholarship.


\end{document}